\documentclass[english]{article}
\usepackage[T1]{fontenc}
\usepackage[latin9]{inputenc}
\usepackage{units}
\usepackage{amsmath}
\usepackage{amssymb}
\usepackage{stmaryrd}

\makeatletter
\newcommand{\lyxaddress}[1]{
	\par {\raggedright #1
	\vspace{1.4em}
	\noindent\par}
}

\makeatother

\usepackage{babel}
\begin{document}
\title{Electron spin correlations: a geometric representation }
\author{Ana María Cetto}
\maketitle

\lyxaddress{Instituto de Física, Universidad Nacional Autónoma de México, Mexico}
\begin{abstract}
An analysis is made within the quantum formalism of the probabilistic
features of the electron spin correlation, with the purpose of clarifying
the concepts of contextuality and measurement dependence. The quantum
formulas for the spin correlation are then derived on the basis of
a probability distribution function and its associated geometric representation,
both for a pair of projections of a single-particle spin and for the
bipartite system in singlet spin state. This endows the spin correlation
with a clear probabilistic meaning and leaves the door open for a
possible physical image of the electron spin, as discussed at the
end of the paper. 
\end{abstract}

\section{Introduction}

The question of whether the mathematical formalism of quantum mechanics
implies necessarily a different kind of probabilities from those used
in classical statistical mechanics, is a subject of continuing debate.
Clarification of the issue is not only of fundamental importance for
a better understanding of quantum theory and a demistification of
the quantum phenomenon, including issues such as nonlocality, acausality
or the absence of realism; it also has important implications for
the development and extension of probability theory with a view to
its applications in other areas, as complex and diverse as epidemiology,
finances, game theory and cognitive science (see, e. g., \cite{Li93},
\cite{AKh10} and references therein).

The present paper is an attempt to contribute to this clarification,
by addressing the question: is there a quantum probability that is
unique in the sense that it does not apply outside of quantum systems?
In other words, is the unusualness of the quantum formalism rooted
in its probabilistic framework, and does it imply the need to renounce
basic principles that hold for the rest of physics? For this purpose
we focus on the electron spin correlation, and make an analysis of
the various probabilistic features provided by the quantum formalism. 

Two conceptual elements that are shown to play a central role in the
present analysis are the context, and the conditional probabilities.
A distinction is made between the notion of context used here to refer
to the measurement that is carried out---i. e., \emph{what} is being
measured---as opposed to the notion of contextuality frequently used
in quantum measurement theory to refer to the result of a measurement
being dependent on which other quantity has been measured. By the
same token, conditional probabilities as discussed here are probabilities
conditioned by the context. Such context-conditioning is connected
with the specific partitioning of the probability space, as has been
shown in previous work \cite{CAP20}.

The electron spin has the advantage of being amenable to a geometric
representation and an associated description in terms of a probability
distribution function involving random variables, or what is usually
called a hidden-variable description. This allows us to reproduce
the probabilistic features \cite{Ce21} and derive the quantum result
for the single spin and the bipartite singlet spin correlation. That
both cases can be dealt with following a similar approach is due to
the use of conditional probabilities in calculating the respective
correlations. Further to endowing the probabilities with a concrete
meaning, the result leaves open the possibility of an understanding
of the physics that underlies the quantum description. A proposal
in this regard is advanced at the end of the present paper, in the
light of recent experimental evidence pointing to a finer dynamics
of the spinning electron, which requires further investigation.

The present paper is structured as follows. Section 2 starts with
the introduction of an algebraic representation of the spin projection
probabilities for the single-spin case, which serves to discuss the
notions of contextuality and conditioned probabilities. This representation
is shown briefly to reproduce the basic probabilistic properties predicted
by the quantum formalism for the electron spin correlation. A central
feature of this algebraic approach is the clear separation of the
context (what is being measured) from the state of the system (in
which it is measured). Section 3 focuses on the bipartite singlet
spin state. The quantum description of the spin correlation is shown
to imply a context-dependent disaggregation of the probability space
into mutually exclusive subspaces. In section 4, a probability distribution
function $\rho(\phi)$ is introduced that reproduces the quantum probabilistic
results. This distribution function is shown in section 5 to be amenable
to a geometric representation that gives meaning to the random variable
$\phi$. The paper concludes with a discussion on the possibility
of a physical image for the electron spin.

\section{The spin-1/2 particle}

\subsection{Analysis of contextuality}

In a recent article \cite{Gr21}, Grangier introduces a ``principle
of contextual quantization'', meaning that \emph{whatever the context,
a measurement on a given system gives one modality among N possible
ones, where the value of N characterizes the system. These N modalities
are mutually exclusive, i. e., only one can be realized at a time}.

Thus for example the projection of an electron spin along an arbitrary
direction $\boldsymbol{a}$ gives either +1 or -1. Since $\pm$ are
the only possible outcomes, $N=2$. Assume first that the result of
the projection along $\boldsymbol{a}$ is $+1$; if the spin is measured
again along $\boldsymbol{a},$ the result +1 is again obtained. If
however the projection is measured along a different direction $\boldsymbol{b}$,
one gets $-1$ with a certain probability. This can be expressed by
means of a $2\times2$ matrix of probabilities that depends on the
two directions $\boldsymbol{a}$ and $\boldsymbol{b}$, the rows referring
to the possible signs of $\boldsymbol{a}$ and the columns to those
of $\boldsymbol{b}$:
\begin{equation}
P(b,a)=\left(\begin{array}{cc}
P_{ab}(+\mid+) & P_{ab}(-\mid+)\\
P_{ab}(+\mid-) & P_{ab}(-\mid-)
\end{array}\right),\label{2-2}
\end{equation}
with $P_{ab}(b\mid a)$ the probability of $b$ conditioned by the
value of $a$. Thus for instance, $P_{ab}(-\mid+)$ is the probability
that, given a $+1$ projection along $\boldsymbol{a}$, the projection
along $\boldsymbol{b}$ is $-1$. Clearly, since the projection along
$\boldsymbol{b}$ must be either $+1$ or $-1$, \begin{subequations}\label{Pab}
\begin{equation}
P_{ab}(+\mid+)+P_{ab}(-\mid+)=1,\label{Pab+}
\end{equation}
 and 
\begin{equation}
P_{ab}(+\mid-)+P_{ab}(-\mid-)=1.\label{Pab-}
\end{equation}
\end{subequations} The probabilities are invariant under an inversion
of the sense of the directions $\boldsymbol{a}$ and $\boldsymbol{b}$
that interchanges all the plus and minus signs \begin{subequations}\label{Pab-1}
\begin{equation}
P_{ab}(+\mid+)=P_{ab}(-\mid-),\label{Pab++}
\end{equation}
and
\begin{equation}
P_{ab}(+\mid-)=P_{ab}(-\mid+).\label{Pab+-}
\end{equation}
\end{subequations} The matrix $P(a.b)$ is therefore symmetric, i.
e., $P(a,b)=P(b,a)$, whence $\boldsymbol{a}$ and $\boldsymbol{b}$
may be freely interchanged. Moreover, it is doubly stochastic \cite{AKh10},
because both the rows and the columns add to 1. 

Notice that the matrix coefficients represent \emph{conditional} probabilities,
the upper ones referring to the (+ or -) projections along $\boldsymbol{b}$
conditioned by the $+1$ projection along $\boldsymbol{a}$ and the
lower ones by the $-1$ projection along $\boldsymbol{a}$. The corresponding
\emph{joint} probabilities are given by expressions of the form \cite{Ko56}
\begin{equation}
P_{ab}(++)=P_{a}(+)P_{ab}(+\mid+),\;P_{ab}(-+)=P_{a}(+)P_{ab}(-\mid+),\label{2-8}
\end{equation}
where $P_{a}(+)$ is the probability of the projection along $\boldsymbol{a}$
being $+1$, and similarly for the lower pair. Thus the conditional
probabilities $P_{ab}(+\mid+)$, $P_{ab}(-\mid+)$ restrict the sample
space to the situation in which the projection along $\boldsymbol{a}$
is $+1$, and similarly for $P_{ab}(-\mid-)$, $P_{ab}(+\mid-)$ .
This will be important for the discussion in Section 5. The total
probability is the sum of the respective joint probabilities; thus
for instance 
\begin{equation}
P_{b}(+)=P_{a}(+)P_{ab}(+\mid+)+P_{a}(-)P_{ab}(+\mid-).\label{2-10}
\end{equation}
Clearly, 
\begin{equation}
P_{a}(+)+P_{a}(-)=1,\;P_{b}(+)+P_{b}(-)=1.\label{2-12}
\end{equation}

The correlation of the projections is given by the formula 
\begin{equation}
C(a,b)=\frac{P_{ab}(++)+P_{ab}(--)-P_{ab}(-+)-P_{ab}(+-)}{P_{ab}(++)+P_{ab}(--)+P_{ab}(-+)+P_{ab}(+-)}.\label{2-14}
\end{equation}
On account of Eqs. (\ref{Pab})-(\ref{2-12}), the sum of the joint
probabilities in the denominator gives $1$, and Eq. (\ref{2-14})
simplifies into

\begin{equation}
C(a,b)=P_{ab}(+\mid+)-P_{ab}(-\mid+)=P_{ab}(-\mid-)-P_{ab}(+\mid-).\label{2-16}
\end{equation}
Notice that, by involving the conditional probabilities only, this
result is independent of the total probabilities $P_{a}(\pm)$, $P_{b}(\pm)$.
This is an important feature of the matrix of probabilities, as it
means that it applies to any joint measurement along $\boldsymbol{a}$
and $\boldsymbol{b}$ as described above, \emph{regardless of the
spin state}, i. e., of the preparation of the spin to be measured.
Briefly, one may say that $P(a,b)$ refers to the contextuality of
the measurements, viz the arrangement of the measuring devices, in
line with the meaning of the term 'context' used in Refs. \cite{AKh10},
\cite{Gr21}. 

\subsection{Spin projection probabilities}

To calculate the conditional probabilities for the single spin case,
we use the standard expressions for the bases of spin state vectors
along two arbitrary directions $\boldsymbol{a}$ and $\boldsymbol{b}$
lying on the same vertical plane and forming angles $\theta_{a}$
and $\theta_{b}$, respectively, with the $z$ axis. In terms of $\vartheta_{a,b}\equiv\theta_{a,b}/2$,

\begin{equation}
\left|+\right\rangle _{a}=\left(\begin{array}{c}
\cos\vartheta_{a}\\
-\sin\vartheta_{a}
\end{array}\right),\;\left|-\right\rangle _{a}=\left(\begin{array}{c}
\sin\vartheta_{a}\\
\cos\vartheta_{a}
\end{array}\right),\label{12}
\end{equation}
and similarly for $\left|\pm\right\rangle _{b}$. This gives, with
$\vartheta_{ba}=\vartheta_{b}-\vartheta_{a},$ \begin{subequations}\label{ba}
\begin{equation}
_{b}\left\langle +\right.\left|+\right\rangle _{a}={}_{b}\left\langle -\right.\left|-\right\rangle _{a}=\cos\vartheta_{ba},\label{14a}
\end{equation}
\begin{equation}
_{b}\left\langle +\right.\left|-\right\rangle _{a}=-{}_{b}\left\langle -\right.\left|+\right\rangle _{a}=-\sin\vartheta_{ba}.\label{14b}
\end{equation}
\end{subequations} The conditional probabilities are therefore given
by \begin{subequations}\label{ba2}

\begin{equation}
P_{ab}(+\mid+)=P_{ab}(-\mid-)=\cos^{2}\vartheta_{ba},\label{16a}
\end{equation}
\begin{equation}
P_{ab}(+\mid-)=P_{ab}(-\mid+)=\sin^{2}\vartheta_{ba},\label{16b}
\end{equation}
\end{subequations} whence Eq. (\ref{2-2}) becomes
\begin{equation}
P(b,a)=\left(\begin{array}{cc}
\cos^{2}\vartheta_{ba} & \sin^{2}\vartheta_{ba}\\
\sin^{2}\vartheta_{ba} & \cos^{2}\vartheta_{ba}
\end{array}\right).\label{18}
\end{equation}
From Eq. (\ref{2-16}) we obtain for the correlation of the spin projections
\[
C_{Q}(\boldsymbol{a},\boldsymbol{b})=\left\langle \psi\right|\left(\hat{\boldsymbol{\sigma}}\cdotp\boldsymbol{b}\right)\left(\hat{\boldsymbol{\sigma}}\cdotp\boldsymbol{a}\right)\left|\psi\right\rangle 
\]
the well-known result for the quantum correlation,
\begin{equation}
C_{Q}(a,b)=\cos^{2}\vartheta_{ba}-\sin^{2}\vartheta_{ba}=\cos\theta_{ba},\label{20}
\end{equation}
regardless of the spin state $\left\vert \psi\right\rangle $. 

\subsection{On the 'quantumness' of spin probabilities }

The mathematical element represented by Eq. (\ref{2-2}), with its
associated properties discussed above, is, according to Grangier \cite{Gr21},
a 'fundamentally quantum idea', because with a couple of simple consistency
arguments it leads to the inevitable conclusion that the only possible
theory is quantum mechanics.

The first consistency argument refers to the sum of the projectors,
which must be equal to 1, as indicated in Eqs. (\ref{Pab}), for any
measurement context. The appeal made in \cite{Gr21} to Gleason's
theorem does not apply to the present case, in which we are dealing
with a two-dimensional Hilbert space (\cite{Gl57,Co85}). It would
seem, therefore, that we need to resort to the Kochen-Specker theorem
\cite{Ko67}, which excludes any non-contextual hidden-variable theory
able to reproduce the quantum results, thus assigning a seal of uniqueness
to quantum probabilities. This points to the relevance of establishing
a clear definition of what is meant by contextual, a point to which
we will return in the following sections. The second consistency argument
in \cite{Gr21} refers to the unitarity of the transformations between
projectors, which is necessary to preserve the mutually exclusive
character of events in each context \cite{Uh62}. That this condition
is satisfied can be proved by associating to the probability matrix
$P(b,a)$ given by (\ref{18}), an orthogonal matrix $F_{ba}$ whose
elements are the square roots of the coefficients of $P(b,a)$ , 
\begin{equation}
F_{ba}=\left(\begin{array}{cc}
\cos\vartheta_{ba} & -\sin\vartheta_{ba}\\
-\sin\vartheta_{ba} & -\cos\vartheta_{ba}
\end{array}\right).\label{22}
\end{equation}

Indeed, a change of measuring context, from $(\boldsymbol{a},\boldsymbol{b})$
to $(\boldsymbol{a},\boldsymbol{c})$, with $\vartheta_{ca}=\vartheta_{c}-\vartheta_{a}=\vartheta_{cb}+\vartheta_{ba}$,
changes $F_{ba}$ into $F_{ca}$ via a unitary transformation,
\begin{equation}
F_{ca}=U_{cb}F_{ba},\label{24}
\end{equation}
with the matrix $U_{cb}$ given by 
\begin{equation}
U_{cb}=\left(\begin{array}{cc}
\cos\vartheta_{cb} & \sin\vartheta_{cb}\\
-\sin\vartheta_{cb} & \cos\vartheta_{cb}
\end{array}\right),\label{26}
\end{equation}
and $U_{cb}U_{cb}^{\dagger}=1$. In terms of Pauli matrices, Eqs.
(\ref{24}) and (\ref{26}) take the form
\begin{equation}
F_{ba}=\cos\vartheta_{ba}\sigma_{z}-\sin\vartheta_{ba}\sigma_{x},\label{28}
\end{equation}
\begin{equation}
U_{cb}=\cos\vartheta_{cb}\mathbb{I}+i\sin\vartheta_{cb}\sigma_{y}.\label{30}
\end{equation}
Notice that when operating on $F_{ba}$, the matrix $U_{cb}$ leaves
the right subindex $a$ unchanged. This can be understood by noting
that $U_{cb}$, being an orthogonal matrix, describes a rotation by
an angle $\theta_{cb}$ around the $\boldsymbol{a}$ axis. Since
\begin{equation}
U_{db}=U_{dc}U_{cb},\label{32}
\end{equation}
successive application of $U$ on $F_{ba}$ gives
\begin{equation}
U_{dc}U_{cb}F_{ba}=U_{dc}F_{ca}=U_{db}F_{ba}=F_{da}.\label{34}
\end{equation}
The same matrix $U$, when operating over a vector basis, transforms
it into a new basis. Take, e. g., the initial basis of state vectors
along $\boldsymbol{b}$, given by Eq. (\ref{12}) (with $a\rightarrow b$),
and apply to them the transformation $U_{cb}$, 
\begin{equation}
U_{cb}\left(\begin{array}{c}
\cos\vartheta_{b}\\
-\sin\vartheta_{b}
\end{array}\right)=\left(\begin{array}{c}
\cos\vartheta_{c}\\
-\sin\vartheta_{c}
\end{array}\right),\;U_{cb}\left(\begin{array}{c}
\sin\vartheta_{b}\\
\cos\vartheta_{b}
\end{array}\right)=\left(\begin{array}{c}
\sin\vartheta_{c}\\
\cos\vartheta_{c}
\end{array}\right).\label{36}
\end{equation}
Therefore, the change of measuring context from $(\boldsymbol{a},\boldsymbol{b})$
to $(\boldsymbol{a},\boldsymbol{c})$ implies also a change of vector
basis, from $\left|\pm\right\rangle _{b}$ to $\left|\pm\right\rangle _{c}$. 

Notice that this transformation does not have any effect on the state
of the system. It does, however, introduce a change in the partitioning
of the probability space, reflected in the coefficients of the probability
matrix (\ref{18}).

\section{The entangled (singlet) bipartite system}

\subsection{Separating the contributions to the spin correlation}

Let us now consider a system made of two $\nicefrac{1}{2}-$spin particles
in the (entangled) singlet state 
\begin{equation}
\left\vert \Psi^{0}\right\rangle =\frac{1}{\sqrt{2}}\left(\left\vert +_{r}\right\rangle \left\vert -_{r}\right\rangle -\left\vert -_{r}\right\rangle \left\vert +_{r}\right\rangle \right),\label{5-2}
\end{equation}
in terms of the standard notation $\left\vert \phi\right\rangle \left\vert \chi\right\rangle =\left\vert \phi\right\rangle \otimes\left\vert \chi\right\rangle ,$
with $\left\vert \phi\right\rangle $ a vector in the Hilbert space
of spin 1, and $\left\vert \chi\right\rangle $ a vector in the Hilbert
space of spin 2. The direction $\boldsymbol{r}$ is arbitrary since
the singlet state is spherically symmetric. The projection of the
spin 1 operator along $\boldsymbol{a}$ is described by $(\hat{\boldsymbol{\sigma}}\cdotp\boldsymbol{a})\otimes\mathbb{I}$,
and the projection of the spin 2 operator along $\boldsymbol{b}$
is described by $\mathbb{I}\otimes(\hat{\boldsymbol{\sigma}}\cdotp\boldsymbol{b})$,
whence the correlation is given by
\begin{equation}
C_{Q}(\boldsymbol{a},\boldsymbol{b})=\left\langle \Psi^{0}\right|\left(\hat{\boldsymbol{\sigma}}\cdotp\boldsymbol{a}\right)\otimes\left(\hat{\boldsymbol{\sigma}}\cdotp\boldsymbol{b}\right)\left|\Psi^{0}\right\rangle .\label{5-4}
\end{equation}
 For the calculation of $C_{Q}$ we make use of the individual spin
state vectors (\ref{12}) to construct an orthonormal basis for the
bipartite system: 
\[
\left|\phi^{1}\right\rangle _{ab}=\left|+\right\rangle _{a}\left|-\right\rangle _{b},\ \ \left|\phi^{2}\right\rangle _{ab}=\left|-\right\rangle _{a}\left|+\right\rangle _{b},
\]
\begin{equation}
\left|\phi^{3}\right\rangle _{ab}=\left|+\right\rangle _{a}\left|+\right\rangle _{b},\ \ \left|\phi^{4}\right\rangle _{ab}=\left|-\right\rangle _{a}\left|-\right\rangle _{b},\label{5-6}
\end{equation}
and write
\begin{equation}
C_{Q}(\boldsymbol{a},\boldsymbol{b})=\left\langle \Psi^{0}\right|(\hat{\boldsymbol{\sigma}}\cdotp\boldsymbol{a})\left(\sum_{k=1}^{4}\left|\phi^{k}\right\rangle _{ab}\left\langle \phi^{k}\right|_{ab}\right)(\hat{\boldsymbol{\sigma}}\cdotp\boldsymbol{b})\left|\Psi^{0}\right\rangle .\label{5-8}
\end{equation}
The operators 
\begin{equation}
\hat{P}^{k}(\boldsymbol{a},\boldsymbol{b})=\left|\phi^{k}\right\rangle _{ab}\left\langle \phi^{k}\right|_{ab}\label{5-10}
\end{equation}
appearing in (\ref{5-8}) are the projection operators in the product
space of the individual spin spaces, $\mathcal{S=\mathcal{S}}_{1}\varotimes\mathcal{S}_{2}$,
with respective eigenvalues $A_{k}$ corresponding to the bipartite
states $\left\vert \phi^{k}\right\rangle _{ab}$ and given according
to (\ref{5-6}) by 
\begin{equation}
A_{1}=A_{2}=-1\equiv A^{-},\ A_{3}=A_{4}=+1\equiv A^{+}.\label{5-12}
\end{equation}
This allows us to rewrite Eq. (\ref{5-8}) in the form
\begin{equation}
C_{Q}(\boldsymbol{a},\boldsymbol{b})=\sum_{k=1}^{4}A_{k}(\boldsymbol{a},\boldsymbol{b})C_{k}(\boldsymbol{a},\boldsymbol{b}),\label{5-14}
\end{equation}
which is the appropriate spectral decomposition of the spin correlation.
In terms of the projection operators (\ref{5-10}), we may write the
spin correlation operator in the form 
\begin{equation}
\hat{C}_{Q}(\boldsymbol{a},\boldsymbol{b})=\sum_{k=1}^{4}A_{k}(\boldsymbol{a},\boldsymbol{b})\hat{P}^{k}(\boldsymbol{a},\boldsymbol{b})\equiv\sum_{k=1}^{4}\hat{C}_{k}(\boldsymbol{a},\boldsymbol{b}),\label{5-16}
\end{equation}
with $A_{k}$ the eigenvalues given by Eqs. (\ref{5-12}). The coefficients
appearing in (\ref{5-14})
\begin{equation}
C_{k}(\boldsymbol{a},\boldsymbol{b})=|\left(\langle\phi^{k}|_{ab}\right)|\Psi^{0}\rangle|^{2},\label{5-18}
\end{equation}
which are the relative weights of the eigenvalues $A_{k}$, are calculated
with the help of Eqs. (\ref{5-4}) and (\ref{5-6}), \begin{subequations}
\label{Ck}
\begin{equation}
C_{1}(\boldsymbol{a},\boldsymbol{b})=C_{2}(\boldsymbol{a},\boldsymbol{b})=\frac{1}{2}\cos^{2}\vartheta_{ba},\label{3-4a}
\end{equation}

\begin{equation}
C_{3}(\boldsymbol{a},\boldsymbol{b})=C_{4}(\boldsymbol{a},\boldsymbol{b})=\frac{1}{2}\sin^{2}\vartheta_{ba}.\label{3-4b}
\end{equation}
\end{subequations} The conditional probabilities are therefore given
in this case by \begin{subequations}\label{ba2-1}

\begin{equation}
P_{ab}(+\mid-)=P_{ab}(-\mid+)=\cos^{2}\vartheta_{ba},\label{16a-1}
\end{equation}
\begin{equation}
P_{ab}(+\mid+)=P_{ab}(-\mid-)=\sin^{2}\vartheta_{ba},\label{16b-1}
\end{equation}
\end{subequations} whence Eq. (\ref{2-2}) becomes
\begin{equation}
P(b,a)=\left(\begin{array}{cc}
\sin^{2}\vartheta_{ba} & \cos^{2}\vartheta_{ba}\\
\cos^{2}\vartheta_{ba} & \sin^{2}\vartheta_{ba}
\end{array}\right).\label{18-1}
\end{equation}
Eqs. (\ref{Ck}) inserted into Eq. (\ref{5-14}) reproduce the quantum
result,
\begin{equation}
C_{Q}(\boldsymbol{a},\boldsymbol{b})=-\cos\theta_{ba}.\label{3-6}
\end{equation}

\subsection{Context-dependent partitioning of the probability space}

It is important to observe that each term in the sum (\ref{5-16})
projects onto one and only one of the four mutually orthogonal subspaces
$\mathcal{U}^{k}(\boldsymbol{a},\boldsymbol{b})$ that add to form
space $\mathcal{S}$ \cite{Hass13},
\begin{equation}
S=\mathcal{U}^{1}\oplus\mathcal{U}^{2}\oplus\mathcal{U}^{3}\oplus\mathcal{U}^{4}.\label{2-29}
\end{equation}
In operational terms (\cite{Busch95}, Ch. 2), this means that the
result of every (joint) measurement falls under one and only one of
these (eigen)subspaces. Each of the coefficients $C_{k}$ is therefore
identified with a probability measure, namely the probability of obtaining
$A_{k}$ as the result of a measurement, in accordance with the Born
rule (\cite{Khren}, Ch. 1). 

Let us now consider the observable $C_{Q}(\boldsymbol{a},\boldsymbol{b'})$
with $\boldsymbol{b'}\neq\boldsymbol{b}$. The corresponding projection
operators are 
\begin{equation}
\hat{P}^{k}(\boldsymbol{a},\boldsymbol{b'})=\left|\phi^{k}\right\rangle _{ab'}\left\langle \phi^{k}\right|_{ab'},\label{2-34}
\end{equation}
where $\left|\phi^{k}\right\rangle _{ab'}$ is defined as in (\ref{5-6})
with $b$ replaced by $b'$. Therefore, instead of the partitioning
of $\mathcal{S}$ given by (\ref{2-29}), the spectral decomposition
involves now the partitioning into four mutually orthogonal subspaces
$\mathcal{U}^{k}(\boldsymbol{a},\boldsymbol{b'})$, such that every
(joint) measurement falls under one and only one of these new subspaces.
In other words, the probability subspaces are specific to the observable
being measured, i. e., to the measurement setting. This assigns an
unambiguous meaning to the term \emph{measurement dependence} that
has been introduced in the context of the Bell-type inequalities (see
e. g. \cite{Ver13}): Contrary to a widespread notion of the term
as implying a (functional) dependence of a set of hidden variables
common to the entire probability space on the measurement setting,
according to the present discussion it refers to the dependence of
the \emph{partitioning of the probability space} on the measurement
setting.

This calculation carried out within the quantum Hilbert-space formalism
\cite{CAP20} confirms that the context must in general be considered
when calculating quantum-mechanical probabilistic quantities. Specifically,
the context---in this case, the directions $\boldsymbol{a}$ and
$\boldsymbol{b}$---is shown to entail the division of the entire
probability space into mutually exclusive, complementary probability
subspaces. The two concepts, measurement dependence and contextual
probabilities, are thus seen to be closely linked.

\section{Probability distribution for the electron spin}

In a recent article \cite{Ce21} a general probability distribution
$\rho(\phi)$ has been proposed for the electron spin projection problem,
which serves to reproduce the conditional probabilities and the correlation
$C(a,b)$, for both the single spin and the bipartite singlet state.
This probability distribution has the form\footnote{The same formula for the distribution, Eq. (\ref{40}), has been previously
obtained by other authors, also within the standard framework of quantum
mechanics; see, e. g., \cite{Oak16}. }
\begin{equation}
\rho(\phi)=\frac{1}{2}\sin\phi,\;0\leq\phi\leq\pi,\label{40}
\end{equation}
with
\begin{equation}
\int_{\Phi}\rho(\phi)d\phi=1.\label{42}
\end{equation}
The partitioning of the probability space $\Phi$ into $\Phi_{ab}^{+}$,
$\Phi_{ab}^{-}$ must be such that, according to Eqs. (\ref{ba2})
in the single-spin case,
\begin{equation}
\int_{\varPhi_{ab}^{+}}\rho(\phi)d\phi=\cos^{2}\vartheta_{ab},\int_{\varPhi_{ab}^{-}}\rho(\phi)d\phi=\sin^{2}\vartheta_{ab}.\label{44}
\end{equation}
With $\rho(\phi)$ given by Eq. (\ref{40}), the subdivision is (recall
that $\vartheta_{ab}=\theta_{ab}/2$) \begin{subequations}\label{Phi}
\begin{equation}
\int_{\varPhi_{ab}^{+}}\rho(\phi)d\phi=\frac{1}{2}\int_{\theta_{ab}}^{\pi}\sin\phi d\phi=\cos^{2}\frac{\theta_{ab}}{2},\label{46a}
\end{equation}
\begin{equation}
\int_{\varPhi_{ab}^{-}}\rho(\phi)d\phi=\frac{1}{2}\int_{0}^{\theta_{ab}}\sin\phi d\phi=\sin^{2}\frac{\theta_{ab}}{2}.\label{46b}
\end{equation}
\end{subequations} The correlation $C(a,b)$ is given accordingly
by 
\begin{equation}
C(a,b)=\left(\int_{\varPhi_{ab}^{+}}-\int_{\varPhi_{ab}^{-}}\right)\rho(\phi)d\phi=\cos\theta_{ab},\label{48}
\end{equation}
in agreement with Eq. (\ref{20}). Equation (\ref{40}) can therefore
be considered to represent a bona fide hidden-variable distribution
for the single electron spin. It is important to keep in mind that
the contextuality resides in the partitioning of the sample space,
not in the outcomes of measurements. In other words, the same function
$\rho(\phi)$ applies to different settings; but the \emph{set} of
values of $\phi$ realized in each case to give either $+1$ or $-1$,
depends on the setting.

In the bipartite case the signs of the spin projection along $b$
are inverted with respect to the single-spin case, so that instead
of Eq. (\ref{44}) we have
\begin{equation}
\int_{\varPhi_{ab}^{-}}\rho(\phi)d\phi=\cos^{2}\vartheta_{ab},\int_{\varPhi_{ab}^{+}}\rho(\phi)d\phi=\sin^{2}\vartheta_{ab},\label{50}
\end{equation}
the corresponding subdivision is \begin{subequations}\label{Phi-1}
\begin{equation}
\int_{\varPhi_{ab}^{-}}\rho(\phi)d\phi=\frac{1}{2}\int_{\theta_{ab}}^{\pi}\sin\phi d\phi,\label{46a-1}
\end{equation}
\begin{equation}
\int_{\varPhi_{ab}^{+}}\rho(\phi)d\phi=\frac{1}{2}\int_{0}^{\theta_{ab}}\sin\phi d\phi,\label{46b-1}
\end{equation}
\end{subequations} and the correlation is given accordingly by $C(a,b)=-\cos\theta_{ab},$

\section{Geometric model for the electron spin}

The form of the probability distribution (\ref{40}), along with the
partitioning of the sample space indicated in Eqs. (\ref{Phi}) and
(\ref{Phi-1}), is suggestive of a geometric representation that can
be explored as a basis for a model for the spinning electron \cite{Ce21}.
We shall discuss the single-spin case, and restrict the analysis to
both vectors $\boldsymbol{a}$ and $\boldsymbol{b}$ lying on the
$xz$ plane for simplicity in the discussion. 

In line with the probabilistic description, we are considering an
element pertaining to an ensemble of realizations. Assume we want
to determine $b$, \emph{given a certain value for} $a$, say $a=+1$.
Take for simplicity the $+z$ axis along $\boldsymbol{a}$. We know
for sure that a second spin projection along $\boldsymbol{a}$ gives
again $a=+1$. In terms of the conditional probabilities introduced
in Section 2, 
\begin{equation}
P_{aa}(+\mid+)=1,\;P_{aa}(+\mid-)=0.\label{52}
\end{equation}
This means that the spin vector must lie in the upper half space (or
northern hemisphere), forming in principle any angle measured on the
$xz$ plane. We propose to identify the variable $\phi$ with that
angle; then $\phi$ lies in the interval $0\leq\phi\leq\pi$, with
the origin of $\phi$ along the $+x$ axis and $\phi$ increasing
counterclockwise. Conversely, given $a=-1$, the spin vector must
lie in the lower half space, forming any angle $\phi$ on the $xz$
plane such that $0\leq\phi\leq\pi$, with the origin of $\phi$ along
the $-x$ axis, i. e., $P_{aa}(-\mid-)=1,\;P_{aa}(-\mid+)=0$. (The
argument is of course reversible, in the sense that if $b$ is given,
the angle variable $\phi$ is measured with reference to the direction
of $\boldsymbol{b}$.)

When $a=+1$, the sign of the projection along the direction $\boldsymbol{b}$
forming an angle $\theta_{ab}$ with the $+z$ axis is $b=+1$ for
any angle $\phi$ on that plane such that $\theta_{ab}\leq\phi\leq\pi$,
whilst it is negative for $0\leq\phi\leq\theta_{ab}$. This gives
a concrete geometrical meaning to Eqs. (\ref{44})-(\ref{48}), and
justifies the partitioning of the probability space into the complementary
subspaces $\Phi_{ab}^{+}(\theta_{ab},\pi)$, $\Phi_{ab}^{-}(0,\theta_{ab})$.
What determines in each individual instance the specific value of
the variable $\phi$, is not known here; $\phi$ may vary at random
between realizations. within the entire interval $(o,\pi).$ What
is the source of such randomness and the mechanism that gives rise
to the distribution function $\rho(\phi)$, is also unknown at this
stage. What is important here is that a probability distribution exists
that reproduces the desired conditional probabilities and correlations,
without additional assumptions.

To make the context dependence more explicit, one may rewrite Eq.
(\ref{48}) as
\[
C(a,b)=\int_{\Phi}b(\phi)\rho(\phi)d\phi=\frac{1}{2}\left(\int_{\theta_{ab}}^{\pi}-\int_{0}^{\theta_{ab}}\right)\sin\phi d\phi
\]
\begin{equation}
=\frac{1}{2}\int_{0}^{\pi}\left[sign\sin(\phi-\theta_{ab})\right]\sin\phi d\phi=\cos\theta_{ab}.\label{54}
\end{equation}
When the direction is changed from $\boldsymbol{b}$ to $\boldsymbol{b'}$,
the geometry changes and the probability space is subdivided accordingly,
so that one gets instead

\begin{equation}
C(a,b')=\frac{1}{2}\int_{0}^{\pi}\left[sign\sin(\phi'-\theta_{ab'})\right]\sin\phi'd\phi'=\cos\theta_{ab\text{'}}.\label{56}
\end{equation}
A prime has been added to the integration variable $\phi$ in Eq.
(\ref{56}) to stress that, although the distribution \emph{function}
$\rho(\phi)$ is the same, its \emph{realization} is independent from
the previous one. This means that the individual results obtained
in one context may not be transferred to the other. 

The observation just made has important implications: it ascribes
an unavoidable random character to the variable $\phi$. If the behavior
of the system were deterministic, one could label every individual
element of the ensemble and assign to it a fixed value of $\phi$,
regardless of which projection (whether along $\boldsymbol{b}$ or
$\boldsymbol{b'}$) is being measured. 

An analogous approach, mutatis mutandis, can be followed in the entangled
bipartite case: the spin projections along $\boldsymbol{a}$ and $\boldsymbol{b}$
are now those of particles 1 and 2, and the probabilities are conditioned
by the outcome of spin 1, spin 2 being in this case \emph{antiparallel}
to spin 1. The corresponding change of sign of $b$ is reflected in
the final outcome, $C(a,b)=-\cos\theta_{ab}$ (cf. Eq. (\ref{54})). 

In the conventional terminology, the conclusion is that the 'hidden'
variable $\phi$ with its associated distribution $\rho(\phi)$ does
not serve to $complete$ the quantum description, since the random
element is still present. It does serve, however, the purpose of offering
a better understanding of the probabilistic features of spin within
the context of standard probability theory, and a geometric explanation
for such features. 

To demonstrate that there is indeed no need to abandon classical probability
has been also the motivation behind different computer simulations
that produce results in violation of Bell-type inequalities (e. g.
\cite{Acc2002,Gill03,Wett09,Mich2014}; see also \cite{Drumm19} and
refs. therein). As indicated in Ref. \cite{Gill03}, 'one should not
try to explain away the strange features of quantum mechanics as some
kind of defect of classical probabilistic thinking, but one should
use classical probabilistic thinking to pinpoint these features'.
The present work offers a contribution in this direction. 

\section{Final comment. A possible physical image of spin}

At this point one may ask whether a physical image of the electron
spin can be made compatible with the geometric representation just
discussed, under the condition that $\rho(\phi)$, with $0\leq\phi\leq\pi$,
represents a distribution of random variables. Such image would have
to be consistent with the physical notion of spin as a dynamical quantity,
with an associated intrinsic angular momentum $\boldsymbol{s}$ of
fixed magnitude and a magnetic moment roughly given by $\boldsymbol{\mu}=(e/m)\boldsymbol{s}$. 

In the presence of a constant magnetic field $\boldsymbol{H}$, a
classical, frictionless magnetic spinning body is known to regularly
precess around the direction of $\boldsymbol{H}$ with constant angular
frequency as a result of the torque exerted by the field (see e. g.
Ref. \cite{Go02}). A similar image has been conventionally associated
with the electron, in which case the frequency of precession or Larmor
frequency is given by $\boldsymbol{\omega}_{L}=(e/m)\boldsymbol{H}$.
Even for intense magnetic fields, this frequency is many orders of
magnitude smaller than the spinning frequency, which according to
Dirac's theory is estimated to be of the order of Compton's frequency,
$\omega_{C}=mc^{2}/\hbar\sim10^{21}$ s$^{-1}.$ 

This crude image does not seem to leave any space for the additional
inclination variable represented by $\phi$ in our geometric model,
and even less for the possible random character of this variable.
However, such picture may change in the light of recent experimental
evidence. Observations made with ferromagnetic materials in the pico-
and femtosecond scales (\cite{Nee21}, see also \cite{Be96}), provide
evidence of a spin dynamics far richer than previously assumed, due
to effects of damping and inertia. This makes the study of the dynamics
also quite more complicated, owing to the nonlinearity of the dynamical
equations, which are impossible to solve analytically. Analysis of
the detailed dynamics of the spinning electron is clearly outside
the scope of the present paper. What is relevant, however, to our
discussion, is the theoretical possibility of spin nutations, similar
to the ones of a spinning top, and their experimentally observed appearance,
at a characteristic frequency $\omega_{N}$ much higher than the usual
Larmor precession, yet much smaller than Compton's frequency. These
apparently intrinsic nutations have been established experimentally
thanks to the use of an intense, transient magnetic field from a superradiant
source of frequency close to 10$^{12}$ s, to which the nutating spin
resonates. The lack of such sources had previously hampered the observation
of this nutation dynamics. 

Take now the geometric model described in the previous section, and
consider the dynamics of the electron spinning around its own axis
plus the spin angular momentum precessing around the direction of
the magnetic field, along the $z$ axis, which was defined as the
direction $\boldsymbol{a}$. If in addition the spin vector is allowed
to nutate, and it does so in a highly complex and irregular manner
due to the nonlinearity of the dynamics, it may in principle scan
the entire range of values of $\phi$, from 0 to $\pi$. As long as
we cannot observe this nutation, because of its extremely high frequency,
the angle $\phi$ remains as a 'hidden variable'. We are not able
to determine the variations of $\phi$ that occur with such high resolution,
we only know that on the average they must be described by a distribution
function such as $\rho(\phi)$. Whether the randomness of $\phi$
is due to the permanent interaction of the spinning electron with
the fluctuating vacuum, or whether it is a product of the chaotic
behavior of spin at this scale, is an open question; in any case,
there is no need to think that randomness is an inherent element of
physics. 

With this discussion we hope to have provided elements in favor of
the plausibility of a physical explanation for the probabilistic description
of the electron spin given by quantum mechanics, thereby avoiding
the need to resort to arguments of an unphysical or spooky nature.
To conclude, we may briefly say that, although the electron spin itself
is a quantum property, whose dynamics is still in need of a more complete
theory, the current probability theory seems well suited for an explanation
of its probabilistic features.

\end{document}